\documentclass[fleqn,usenatbib]{mnras}

\usepackage{newtxtext,newtxmath}

\usepackage[T1]{fontenc}
\usepackage{ae,aecompl}

\usepackage{graphicx}	
\usepackage{amsmath}	
\usepackage{amssymb}	

\title[Minimum mass of NS]{On the minimum mass of neutron stars}

\author[Y. Suwa et al.]{
Yudai Suwa,$^{1,2}$\thanks{E-mail: suwa@yukawa.kyoto-u.ac.jp}
Takashi Yoshida,$^{3}$\thanks{E-mail: tyoshida@astron.s.u-tokyo.ac.jp}
Masaru Shibata,$^{4,1}$
Hideyuki Umeda,$^{3}$
\newauthor
and Koh Takahashi$^{5}$
\\
$^{1}$Center for Gravitational Physics, Yukawa Institute for Theoretical Physics, Kyoto University, Kyoto 606-8502, Japan\\
$^{2}$Department of Astrophysics and Atmospheric Sciences, Faculty of Science, Kyoto Sangyo University, Kyoto 603-8555, Japan\\
$^{3}$Department of Astronomy, Graduate School of Science, University of Tokyo, Tokyo 113-0033, Japan\\
$^{4}$Max Planck Institute for Gravitational Physics (Albert Einstein Institute), Am M\"uhlenberg 1, Potsdam-Golm, 14476, Germany\\
$^{5}$Argelander-Institute f\"ur Astronomie, Universita\"ut Bonn, D-53121 Bonn, Germany
\\
}

\date{Accepted 2018 September 5. Received 2018 September 4; in original form 2018 August 7}

\pubyear{2018}

\begin{document}
\label{firstpage}
\pagerange{\pageref{firstpage}--\pageref{lastpage}}
\maketitle

\begin{abstract}
We investigate remnant neutron star masses (in particular, the minimum allowed mass) by performing advanced stellar evolution calculations and neutrino-radiation hydrodynamics simulations for core-collapse supernova explosions. We find that, based on standard astrophysical scenarios, low-mass carbon-oxygen cores can have sufficiently massive iron cores that eventually collapse, explode as supernovae, and give rise to remnant neutron stars that have a minimum mass of 1.17 M$_\odot$ --- compatible with the lowest mass of the neutron star precisely measured in a binary system of PSR J0453+1559.
\end{abstract}




\section{Introduction}

The mass of neutron stars (NSs) is one of the most important observables to probe high-density nuclear physics. In particular, the maximum mass of NSs gives a stringent constraint on the nuclear physics above the saturation density. The current largest mass is $\approx$ 2 M$_\odot$ \citep{demo10,anto13}, which means that hypothetical nuclear equations of state having maximum NS mass smaller than 2 M$_\odot$ are excluded. 

Neutron star masses have a broad distribution. A precise measurement is possible for a binary system which contains at least one pulsar \citep{ozel16}. Recently, the first asymmetric system of double neutron stars, PSR J0453+1559, was discovered \citep{mart15}. The secondary NS's mass is much smaller than a canonical mass, that is, $1.174\pm0.004$ M$_\odot$. A corresponding baryonic mass 
of this NS is $\approx$ 1.28 M$_\odot$, which is remarkably smaller than a
typical mass of Fe cores, i.e. $\sim$1.3--1.6 M$_\odot$ \citep[e.g.,][]{sukh18}.
From this observation, a natural question arises: {\it Is it possible to form such a low-mass NS within the standard scenario of supernova (SN) explosion?} 

Although the explosion mechanism of core-collapse SNe is still unclear, there is the standard paradigm of neutrino-driven explosion \citep{beth85}, in which neutrinos produced in the vicinity of newly-born NSs heat up the post-shock material. Aided by the multi-dimensional hydrodynamic effects, e.g. convection and standing accretion shock instability, explosions driven by neutrino heating have been reported in the past decade \cite[see][for recent reviews and references therein]{burr13,jank16}.\footnote{
Note that the current simulations do not account for the observed explosion energy because they are not evolved for a sufficiently long time to see the convergence of the explosion energy. In addition, the nickel amount would be a more difficult to be explained by the current simulations because of the small growth rate of the explosion energy \citep{suwa17}.} In this work, we assume that SN explosion is driven by the neutrino heating mechanism.

Producing a low-mass NS is also an issue for massive star evolution.
Evolved stars with a carbon-oxygen (CO) core heavier than $\sim$1.37 M$_\odot$ have a possibility of SN explosion and NS formation \cite[e.g.][]{nomo87}.
Single stars with a zero-age-main-sequence (ZAMS) mass of $\sim$8--12 M$_\odot$ have a path to electron-capture (EC) SNe \citep{nomo87,taka13} or core-collapse SNe from a low-mass Fe core \citep[e.g.][]{woos80,nomo84,nomo88,umed12,woos15}.
Studies on a large number of progenitors suggest that these stars would produce low-mass NSs \citep{ugli12,ertl16,sukh16}. Note that fall-back accretion would increase NS mass for stars with a massive envelope, and thus, the formation mechanism of the low-mass NS is not trivial.

In close-binary systems, an ultra-stripped SN is a possible path to produce a low-mass NS \citep[e.g.][]{taur13,taur15,suwa15,mori17,mull18}.
Most of the H and He envelopes for these stars could be lost during their evolution by the close binary interactions.
The explosion process of ultra-stripped SNe, especially the mass accretion history onto the proto-neutron star (PNS), would be considerably different from those of single stars.
Thus, in this paper, we investigate NS formation in ultra-stripped SNe.

In the following, we investigate the path to produce such a low-mass NS based on standard methods of stellar evolution and supernova explosion simulations.
As in the previous works \citep{suwa15,yosh17}, which explored the evolution of massive stars whose envelope is supposed to be significantly stripped, we systematically study the dependence of the stellar evolution on the initial CO core mass in a parametric manner. 
By extending previous work toward lower CO core mass, we evaluate the minimum mass of an NS determined from stellar evolution and demonstrate that a low-mass NS like PSR J0453+1559 can be produced in the standard scenario of binary NS formation.
The paper is organized as follows. Section \ref{sec:evolution} describes our stellar evolutionary calculations, in particular focusing on the consequent core masses. 
Section \ref{sec:coremass} gives estimates of Chandrasekhar mass which depends on the profiles of the electron fraction and entropy as well as the iron core mass.
The numerical method of subsequent radiation hydrodynamics simulations and the results are presented in Section \ref{sec:explosion}. 
In Section \ref{sec:ECSN} we discuss differences of EC SNe from Fe-core forming core-collapse SNe in ultra-stripped SNe.
We summarize our results in Section \ref{sec:summary}.

\section{CO core models}
\label{sec:evolution}

We calculate the evolution of nine CO cores with masses 1.35--1.45 M$_\odot$ from the central C burning.
We denote the CO core model with the mass of x.yz M$_\odot$ as COxyz model.
These CO cores correspond to the hypothetically secondary star in a close binary system, which is supposed to lose their H and He-rich envelope during the binary evolution with a primary NS.
The evolution calculations are conducted with the stellar evolution code used in \citet{suwa15} (see also \citealt{yosh17}).
We include the Coulomb corrections for the weak interaction rates which depend on the temperature and the electron number density \citep{toki13}.
We take about 1000 mass zones in each model.

The nuclear reaction network of 300 species of nuclei is adopted.
The adopted nuclear species are as follows; $^1$n, $^{1-3}$H, $^{3,4}$He, $^{6,7}$Li, $^{7,9}$Be, $^{8,10,11}$B, $^{11-16}$C, $^{13-18}$N, $^{14-20}$O, $^{17-22}$F, $^{18-24}$Ne, $^{21-26}$Na, $^{22-28}$Mg, $^{27-32}$Si, $^{27-34}$P, $^{30-37}$S, $^{32-38}$Cl, $^{34-43}$Ar, $^{36-45}$K, $^{38-48}$Ca, $^{40-49}$Sc, $^{42-51}$Ti, $^{44-53}$V, $^{46-55}$Cr, $^{48-57}$Mn, $^{50-61}$Fe, $^{51-62}$Co, $^{54-66}$Ni, $^{54-66}$Cu, $^{59-71}$Zn, $^{61-71}$Ga, $^{63-75}$Ge, $^{65-76}$As, $^{67-77}$Se, $^{70-79}$Br. The isomeric state of $^{26}$Al is taken into account.
We set the initial chemical compositions of the CO cores as evaluated in \citet{suwa15}.
The initial chemical compositions of the CO cores are evaluated using the evolution of H and He burnings.
The mass fractions of C and O for these models are assumed to be 0.360 and 0.611, respectively.
We have confirmed that properties of ultra-stripped SN progenitors do not strongly depend on the C/O ratio in \citet{suwa15}.
Detailed evolution properties are described in Appendix \ref{sec:app1}.

\begin{table*}
\centering
\caption{Model summary.}
\label{tab:model}
\begin{tabular}{l|cccccccc}
\hline
Model & $M_{\rm CO}$ & $M_{\rm ZAMS}$ & $M_{s=3}$ & $M_{Y_e=0.495}$ & $M_{\rm Fe}$ & $M_\mathrm{NS, bary}$ & $M_\mathrm{NS, grav}$\\
 & (M$_\odot$) & (M$_\odot$) & (M$_\odot$) & (M$_\odot$) & (M$_\odot$) & (M$_\odot$) & (M$_\odot$) \\

\hline
CO137  &     1.37 &     9.35 &    1.347 &    1.314 &    1.280 &    1.289 &    1.174 \\
CO138  &     1.38 &     9.40 &    1.349 &    1.316 &    1.274 &    1.296 &    1.179 \\
CO139  &     1.39 &     9.45 &    1.350 &    1.320 &    1.258 &    1.302 &    1.184 \\
CO140  &     1.40 &     9.50 &    1.305 &    1.302 &    1.296 &    1.298 &    1.181 \\
CO142  &     1.42 &     9.60 &    1.284 &    1.280 &    1.265 &    1.287 &    1.172 \\
CO144  &     1.44 &     9.70 &    1.275 &    1.219 &    1.234 &    1.319 &    1.198 \\
CO145  &     1.45 &     9.75 &    1.362 &    1.270 &    1.277 &    1.376 &    1.245 \\
\hline
\end{tabular}
\end{table*}

We evaluate the mass range of single star ZAMS that form a CO core with 1.35--1.45 M$_\odot$ is 9.25--9.75 M$_\odot$ using the same manner in \citet{suwa15}.
The evolution of stars that have the similar range of CO core was investigated in \citet{woos15}.
We compare the mass range of our study with their result.
From Table 1 of \citet{woos15}, the ZAMS mass range of single stars that form a CO core with 1.35--1.45 M$_\odot$ is about 8.8--9.3 M$_\odot$.
The relation between the total ZAMS mass and the CO core mass depends mainly on the overshoot treatment of convective layers.
Strong overshoot gives a large CO core mass for a given ZAMS mass.
Thus, the overshoot effect during the H and He burnings in our models would be weaker than their single star models.
The overshoot is actually constrained from observations of main-sequence stars. 
We use the overshoot parameter with the description in \citet{taka14} until the termination of the He-core burning and the parameter value of $f_{\rm ov} = 0.015$, which reproduces the main-sequence band width observed for AB type stars in open clusters in the Galaxy \citep{maed89}.
The overshoot parameter also has an uncertainty by observations and input physics of stellar evolution models.
The ZAMS mass range of single stars that form a CO core with the above mass range would have an uncertainty of about 1 M$_\odot$ due to the uncertainty of the overshoot parameter.

We obtain a critical mass for Ne ignition in the CO core models through the evolution calculations.
The CO core models with $M_{\rm CO} \ge 1.36$ M$_\odot$ start with off-center Ne burning.
The evolution of the CO cores with $M_{\rm CO} \ge 1.37$ M$_\odot$ is calculated until the central density reaches $\sim$10$^{10}$ g cm$^{-3}$.
These stars form an Fe core.
For the CO136 model, the calculation is stopped when the density becomes 10$^{9.5}$ g cm$^{-3}$ and the Fe layer is formed.
Although we did not calculate further evolution, we expect that this model will form an Fe core and will collapse.
On the other hand, the CO135 model does not cause Ne ignition.
The temperature rises to $1.2 \times 10^9$ K at the mass coordinate of 0.97 M$_\odot$ after the C shell burning and the  temperature starts to decrease.
We continue the calculation until the central density becomes $10^{9.1}$ g cm$^{-3}$ and the central temperature becomes below $10^{8.15}$ K.
Thus, a critical mass for Ne ignition is $\sim$1.36 M$_\odot$ in our CO core models.
\citet{nomo84} showed a critical mass of 1.37 M$_\odot$ for Ne ignition from the evolution calculation of pure Ne star models.
Recently, \citet{schw16} showed a critical mass of 1.35 M$_\odot$ for Ne ignition of pure Ne star models using stellar evolution code MESA \citep[e.g.][]{paxt15}.
Although our model is not an Ne star model and some input physics are different, the critical mass of CO star models for Ne ignition in our study is close to the criteria in these previous studies.

The main properties of the CO core models with $M_{\rm CO} \ge 1.37$ M$_\odot$ are listed in Table \ref{tab:model}.
The Fe core mass at the last moment of the stellar evolution is evaluated using three different criteria: the entropy ($s \le 3$ in units of $k_B$ per baryon, where $k_B$ is Boltzmann's constant),\footnote{Although for canonical SN progenitors $s=4$ is used to determine the position of O-Si layer \citep[e.g.,][]{ertl16}, we here use $s=3$ since our progenitor models exhibit a lower entropy than canonical progenitor models. } the electron fraction ($Y_e \le 0.495$), and the chemical composition ($X$(^^ ^^ Fe") $>$ $X$(^^ ^^ Si"), where $X$(^^ ^^ Fe") and $X$(^^ ^^ Si") denote the mass fractions of Fe-peak elements ($Z \ge 22$) and intermediate elements ($14 \le Z \le 21$) with $Z$ being atomic number). 
We find that the Fe core mass is about $\sim$1.3 M$_\odot$ in these models.
This mass is roughly determined by the arrival position of the convection during the O-shell burning after the off-center O burning.
More details are described in the next section.

\section{Core masses}
\label{sec:coremass}

In this section, we discuss the evolution of the core and its critical mass, above which the core becomes unstable against the self-gravity, to investigate the condition of core collapse.

\begin{figure}
\centering
\includegraphics[width=.45\textwidth]{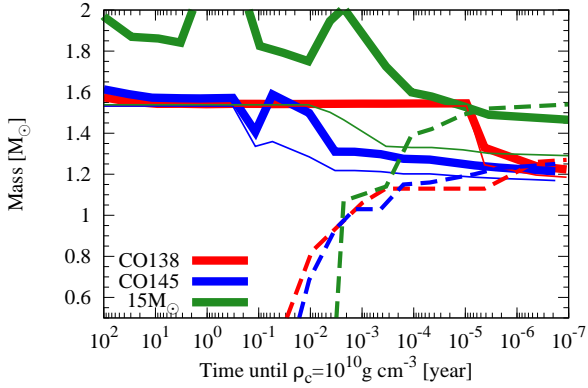}
\caption{Time evolution of modified Chandrasekhar mass (see Eq. (\ref{eq:MCh})) (solid curves) and Fe core mass (dashed curves), for models CO138 (red), CO145 (blue), and a model with $M_{\rm ZAMS}$ = 15 M$_\odot$ for comparison. In the thin solid curves, the thermal correction is not taken into account, while the thick solid curves include the correction. The green dot-dashed line represents the modified Chandrasekhar mass with a constant $Y_e$ structure (see the text for detail).
}
\label{fig:t-M}
\end{figure}

Figure \ref{fig:t-M} shows time evolution of {\it modified} Chandrasekhar mass \citep{baro90,timm96} and the Fe core mass. In the modified Chandrasekhar mass, the finite temperature correction is taken into account as follows,
\begin{equation}
M_{\rm Ch}=M_{\rm Ch0}\left[1+\left(\frac{s_e}{\pi Y_e}\right)^2\right],
\end{equation}
where $M_{\rm Ch0}=1.46$M$_\odot(Y_e/0.5)^2$ is the Chandrasekhar mass without finite temperature correction and $s_e$ is the electronic entropy per baryon. The entropy is typically given as $s_e=0.56(Y_e/0.5)^{2/3}(T/1{\rm\,MeV})(\rho/10^{10}{\rm cm^{-3}})^{-1/3}$ $k_B$ baryon$^{-1}$ with $\rho$ and $T$ being the density and temperature \citep{baro90}. 
In this paper, we derive an expression in which $Y_e$ distribution is taken into account (see Appendix \ref{sec:app2}) as
\begin{equation}
M_{\rm Ch}=
1.09M_\odot
\left(\frac{Y_{e,c}}{0.42}\right)^2
\left[1+\left(\frac{s_{e,c}}{\pi Y_{e,c}}\right)^2\right],
\label{eq:MCh}
\end{equation}
where $Y_{e,c}$ and $s_{e,c}$ are the central values of $Y_e$ and $s_e$.
Figure \ref{fig:t-M} shows that $M_{\rm Ch}$ decreases with time just prior to the onset of collapse because the electron capture and neutrino cooling reduces $Y_e$ and entropy, respectively. The Fe core mass (the outermost mass coordinate that has an Fe mass fraction larger than 0.5), on the other hand, increases by shell burning of Si. 
The collapse takes place when its mass increases over $M_{\rm Ch}$. If we neglect the finite temperature correction, the core mass exceeds $M_{\rm Ch}$ much earlier. We also plot a canonical single star evolution (for $M_{\rm ZAMS}=15$ M$_\odot$) in this figure, which has a larger correction of the finite temperature effect than the present CO core models so that it has a larger core mass when it collapses.
To see the dependence of $Y_e$ distribution on the modified Chandrasekhar mass, two lines of the modified Chandrasekhar mass are shown for 15M$_\odot$ model. The structured $Y_e$ model ($N=3.3$) is the green solid line, while the constant $Y_e$ model ($N=3$) is the green dot-dashed line (see Appendix \ref{sec:app2}). It is easily seen that the structured $Y_e$ model is more compatible to the Fe core evolution and collapse onset. 

\begin{figure*}
\centering
\includegraphics[width=.6\textwidth]{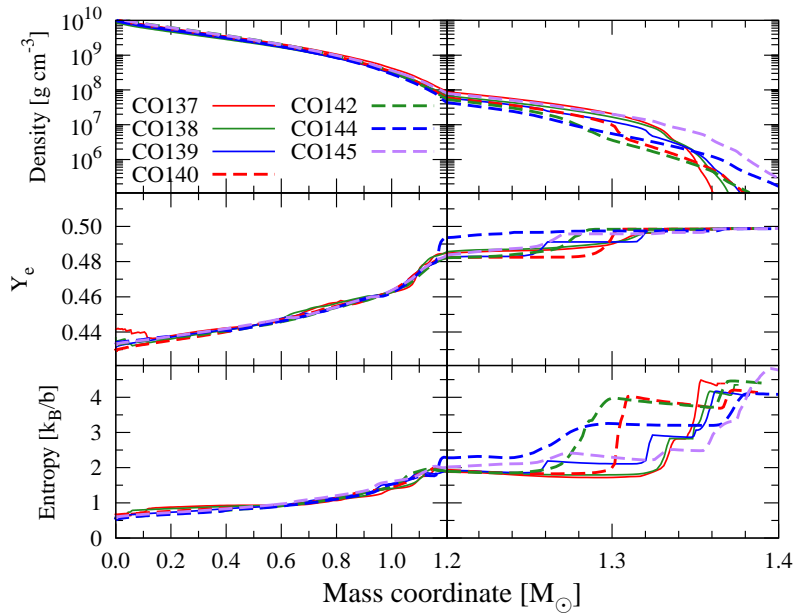}
\caption{Density (top), electron fraction (middle), and entropy (bottom) structure as functions of mass coordinate for investigated models.}
\label{fig:structure}
\end{figure*}

In Figure \ref{fig:structure}, the density, the electron fraction, and the entropy profile when the central density becomes $10^{10}$ g cm$^{-3}$ (approximately at the onset of core collapse) are presented. The central part ($M\lesssim 1.2$ M$_\odot$) is rather similar among all the models shown here, but the outer part depends strongly on the value of $M_{\rm CO}$.
This difference stems primarily from the extension of strong O-shell burning after the off-center O burning.
The decrease in the electron fraction of this region is occurred by the O-shell burning.
In the models with $M_{\rm CO}<1.44$ M$_\odot$, the convective region of the O-shell burning extends to $\sim$1.2 M$_\odot$ or more (see panels ($e$) and ($f$) of Fig. \ref{fig:mf142co} for CO142 model).
On the other hand, the convection does not reach $\sim$1.2 M$_\odot$ for CO144 and CO145 models.
The electron fraction for the mass coordinate in the range of 1.20--1.27 M$_\odot$ for CO145 model is reduced by the following O-shell burning.
For CO144 model, the off-center Si burning starts just after the O-shell burning sets in followed by the off-center O burning.
The region outside the convective Si layer expands and the O-shell burning is ceased.
Thus, the electron fraction is not reduced for $M_r \ga 1.2$ M$_\odot$ for CO144 model.

\section{Explosion simulations}
\label{sec:explosion}

Starting from the progenitor models described in the previous section, we perform two-dimensional neutrino-radiation hydrodynamics simulations, similar to \cite{suwa15,yosh17}. In the previous works, we employed slightly heavier CO cores from 1.45 to 2 M$_\odot$ to make comparison with \cite{taur13}, in which a star with 1.5 M$_\odot$ was investigated to account for a candidate of the ultra-stripped supernova SN 2005ek. This work, on the other hand, is more interested in the final mass of one NS so progenitor models with a lower mass range are explored.

The numerical method is the same as \cite{suwa15}, in which two-dimensional hydrodynamics equations are solved as well as neutrino-radiation transfer equation with isotropic diffusion source approximation (IDSA) \citep{lieb09,suwa10}. Ray-by-ray plus approximation \citep{bura06a} is used to treat the multi-dimensional transfer with a spherically symmetric solver. The nuclear equation of state from \cite{latt91} with incompressibility parameter $K=220$ MeV is employed. 
Although the general relativistic correction for the gravitational potential is not taken into account, the results would not be affected very much because of the steep density gradient in the vicinity of the core surface, which leads to an early explosion onset.

The hydrodynamic properties are the same as those found in \cite{suwa15}. After the onset of core collapse, it takes $\mathcal{O}(100)$ ms until an NS forms, depending on progenitor models (403, 714, 1164, 1022, 1399, 807, and 698 ms for CO137, CO138, CO139, CO140, CO142, CO144, and CO145, respectively). 
They all result in explosions aided by convection in two-dimensional simulations. 
The explosion sets in when the density jump between the interface of Si and Si/O layers accretes onto the shock and the ram pressure above the shock decreases much faster than the thermal pressure below the shock \citep[see e.g.,][]{suwa16,summ16}. Therefore, in the zeroth order approximation, the NS mass is determined by the mass coordinate of the interface \citep[see also][]{sukh18}.

The consequent NS masses are shown in Table \ref{tab:model}. We calculate the gravitational mass of NSs from the baryonic mass with Eq. (35) of \cite{latt01}.\footnote{The mass decrease due to the binding energy of an NS is given as $\Delta M=M_{\rm NS,bary}-M_{\rm NS,grav}=0.084$ M$_\odot(M_{\rm NS,grav}/$M$_\odot)^2$ in this equation. The numerical factor (0.84 in the present case)  depends on the nuclear equation of state, roughly 0.06--0.1 for the mass range we are interested in (see Figure 8 of \citealt{latt01}). This means that for $M_{\rm NS,grav}=1.17$ M$_\odot$, the corresponding baryonic mass is 1.25--1.31 M$_\odot$. Note that the equation of state used in this study \citep{latt91} would give $\sim$0.06--0.08, but we use 0.084. This is because the nuclear equation of state has yet a large uncertainty and the number depends not only on the mass but also on the central density so that a single number for the conversion between the baryonic mass and the gravitational mass has an ambiguity. Thus, we use a canonical number, 0.084. If we take 0.1, which is the maximum value in \cite{latt01}, the minimum gravitational mass of NS would become 1.15M$_\odot$, which will be tested by the future pulsar searches.} Note that the observable in binary pulsar systems is the gravitational mass. It is seen from this table that the gravitational mass of an NS can be as small as 1.17 M$_\odot$, which is consistent with the current observation of the smallest NS mass measured precisely.

\section{Comparison with electron-capture supernovae}
\label{sec:ECSN}

The evolution of massive stars toward EC SN in the binary system with an NS has been discussed in \citet{taur15}.
We suppose that an ultra-stripped EC SN may occur for a CO core with the mass less than 1.36 M$_\odot$ just after the C burning.
An EC SN occurs when a CO core reaches the Chandrasekhar mass by the CO-core growth.
In this case, it is important to clarify how the CO-core grows through the He-shell burning.
In the case of a single star, the ONe-core mass of an EC SN progenitor is less than $\sim$1.37 M$_\odot$ after the C burning \citep[e.g.][]{nomo87,taka13}.
The ONe core mass increases to the modified Chandrasekhar mass through the He shell burning during the evolution of the super asymptotic giant branch (AGB) star.
In an He star in the binary system, there is no H-rich envelope and, thus, AGB phase does not occur.
For such a situation, the CO core growth depends on the efficiencies of the He shell burning and the mass loss of the He envelope.
Because the mass loss process in the binary system is not fully understood and long-time calculation of the CO-core growth through the He-shell burning is required, it is not easy to evaluate whether a CO core less than 1.36 M$_\odot$ in the binary system grows up to the critical mass and explodes as an ultra-stripped EC SN.
The mass range of CO cores for EC SNe was recently discussed for the primary stars in binary system \citep{poel17,sies18}.
However, the mass range would be still model dependent and, thus, we are not sure whether the binary evolution affects the mass range of CO cores of EC SN progenitors.
The investigation of the growth of the ONe core with thin He layer is important to clarify the possibility of the evolutionary path to an EC SN in close binary systems.

How is low-mass NS formed from EC SN progenitors and ultra-stripped SN progenitors that have a low-mass Fe core?
In the case of EC SNe, the electron fraction in the ONe core is $\sim$0.48 at the central Ne ignition \citep{taka13,taka18}.
The electron fraction in the ONe core scarcely decreases until the central Ne ignition.
On the other hand, progenitors of ultra-stripped SNe cause the electron capture during their evolution and, as a result, the electron fraction of the Fe core decreases.
The electron fraction at the center is $\sim$0.43 when the central density becomes $\sim$10$^{10}$ g cm$^{-3}$ in the collapsed models in this paper.
Thus, the Chandrasekhar mass of EC SN progenitors is heavier than that of the progenitors that have a low-mass Fe core of ultra-stripped SNe.
The hydrodynamical simulation in our previous study \citep{yosh17} showed that the baryon mass of the PNS of an EC SN is 1.32 M$_\odot$.\footnote{It should be noted that the remnant NS's mass depends on the numerical methods, ranging from 1.294 to 1.363 $M_\odot$ for the baryonic mass \citep{kita06,jank08,fisc10,radi17}, which correspond to $\sim$1.18 to 1.24M$_\odot$ in the gravitational mass.}
This mass is higher than the baryonic mass range of the NSs formed from CO137--CO144 models.

\section{Summary}
\label{sec:summary}

In this paper, we investigated 
the minimum mass of NSs based on astrophysical scenarios, i.e., stellar collapse and supernova explosion. We calculated the evolution of CO cores with masses 1.35--1.45 M$_\odot$.
The stars with CO-core masses higher than 1.36 M$_\odot$ form an Fe core after off-center Ne, O, and Si-burnings. We found that low-mass CO cores, which eventually form an Fe core and subsequently collapse, could result in an NS with mass $\sim 1.17$ M$_\odot$, which is comparable with the lowest NS mass precisely measured.

\begin{figure}
\centering
\includegraphics[width=.45\textwidth]{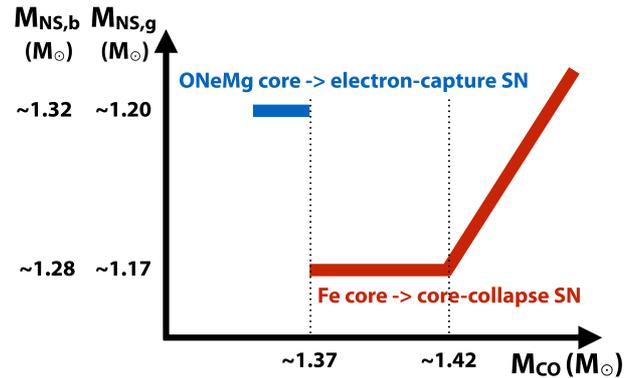}
\caption{A schematic relation between NS mass and CO core mass. Two numbers in NS mass are baryonic mass ($M_{\rm NS,b}$; left) and gravitational mass ($M_{\rm NS,g}$; right), respectively. For $M_{\rm CO}\lesssim $1.37M$_\odot$, the electron-capture SN would be produced and the consequent NS mass would be $\sim$1.37M$_\odot$ and 1.24M$_\odot$ in baryonic mass and gravitational mass. For $M_{\rm CO}\gtrsim$ 1.37 M$_\odot$, the core-collapse SN would be produced and because of its lower-mass Fe core the consequent mass of the NS is smaller than models which produce EC SN.}
\label{fig:mco2mns}
\end{figure}

According to our stellar evolution simulations, the minimum mass of CO cores that produce an Fe core is $\sim 1.37$M$_\odot$ and below this value a ONe core is formed, which would lead to EC SN instead. Because of its higher value of $Y_e$, an EC SN would produce more massive NSs than a core-collapse SN from an Fe core. Therefore, the minimum mass of NSs is expected to be determined by the core-collapse SN of a low-mass CO core (see Figure \ref{fig:mco2mns}).

The range of the Fe core mass for single stars of initial mass 9--10 M$_\odot$ in \citet{woos15} is similar, or smaller, than ultra-stripped SN progenitors.
Thus, the lowest-mass NSs may also be formed from the collapse of low-mass Fe cores.
In the case of single stars, however, the reverse shock is produced when the shock wave arrives at the interface of the H-rich envelope.
Then, the fall-back material accretes onto the collapsed core and increases NS mass.
In the case of ultra-stripped SNe, on the other hand, the fall-back will be negligible because of very thin He layer and no H-rich envelope.
Due to the same reason, the primary NS generated by the first SN explosion (not ultra-stripped SN) in binary systems would be incompatible to the light NS in PSR J0453+1559. This is because the secondary star would supply mass to the NS during its giant phase and increase the NS mass.
Thus, low-mass Fe cores in the progenitors of ultra-stripped SNe, which are conjectured second explosions in close-binary systems, would be more favorable to form lowest-mass NSs.

\section*{Acknowledgements}

YS thanks J. M. Fedrow for proofreading.
TY thanks T. M. Tauris, N. Langer, and T. J. Moriya for discussion on the advanced evolution of light CO core and a hospitality during short stays in Argelander-Institute f\"ur Astronomie, Universit\"at Bonn in 2015 and 2018.The numerical computations in this study were partly carried out on XC30 at CfCA in NAOJ and XC40 at YITP in Kyoto University.
This work was supported in part by the Grant-in-Aid for Scientific Research (Nos. 16H00869, 16H02183, 16K17665, 17H01130, 17H02864, 17H06364, 18H04586, 18H05437, 26400271, 26104007), MEXT as ``Priority Issue on Post-K computer'' (Elucidation of the Fundamental Laws and Evolution of the Universe) and a post K project (No. 9) of the Japanese MEXT.
KT is supported by Japan Society for the Promotion of Science (JSPS) Overseas Research Fellowships.




\bibliographystyle{mnras}
\bibliography{ns_minimum} 




\appendix

\section{Evolution of CO cores}
\label{sec:app1}

\begin{figure}
\centering
\includegraphics[width=.32\textwidth,angle=270]{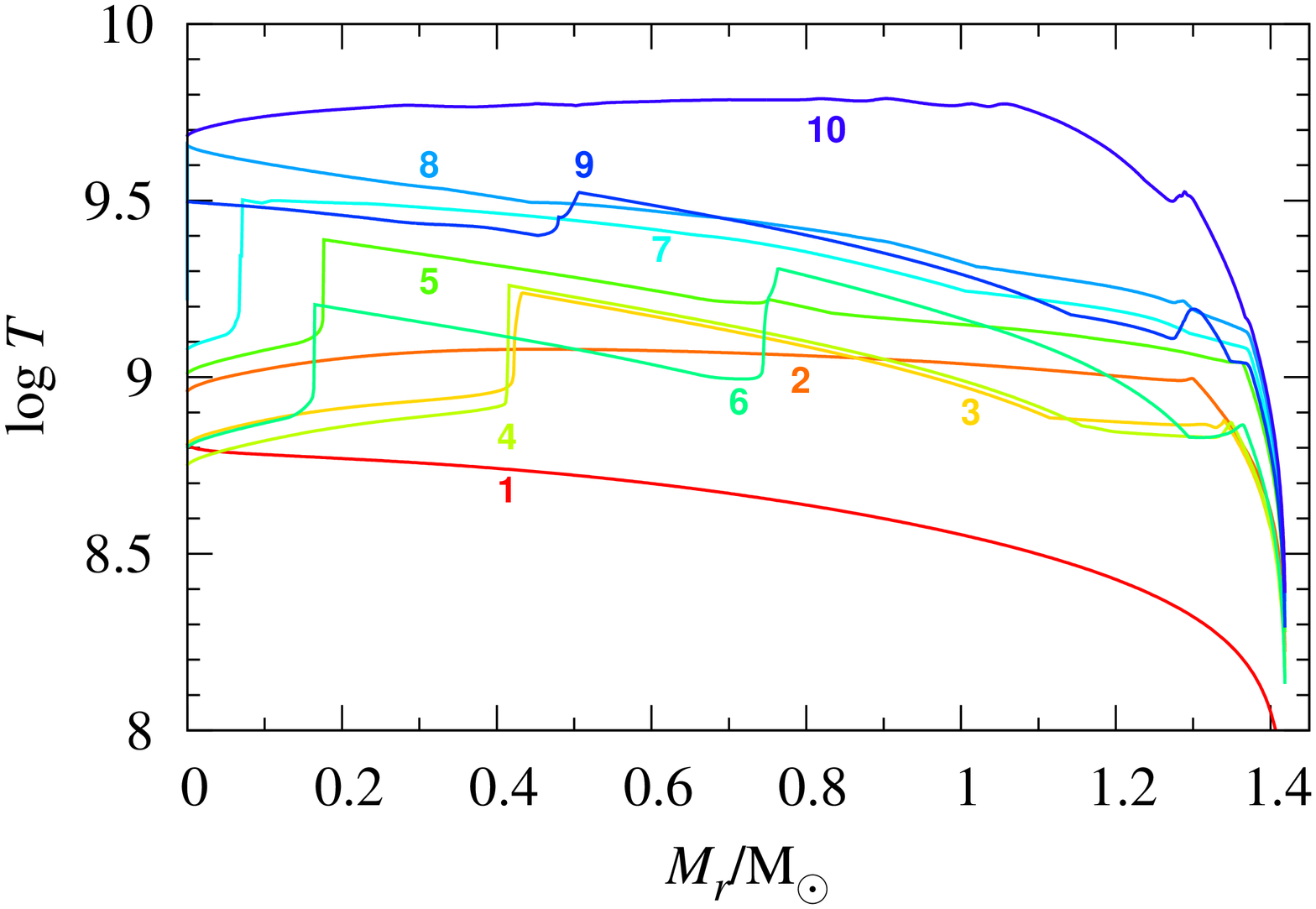}
\caption{Time evolution of the temperature profile of CO142 model.
The evolution numbers correspond to the following: 
(1) the central C burning at $t_f - t = 1.0 \times 10^4$ yr,
(2) the formation of ONe core at $t_f -t = 31.7$ yr, 
(3) off-center Ne burning at $t_f -t = 23.5$ yr, 
(4) off-center O burning at $t_f - t = 8.9$ yr, 
(5) off-center O burning and the growth of the Si layer $t_f - t = 1.6$ yr, 
(6) O-shell burning at $t_f - t = 1.4$ yr, 
(7) off-center Si burning and Fe layer formation $t_f - t = 4.7$ d, 
(8) the flame arrival at the center and the Fe-core formation at $t_f - t = 2.8$ d, 
(9) Si-shell burning and Fe core growth at $t_f - t = 2.4$ d, 
(10) the last profile ($t = t_f$).}
\label{fig:temp1p42co}
\end{figure}

\begin{figure*}
\centering
\includegraphics[width=.2\textwidth,angle=270]{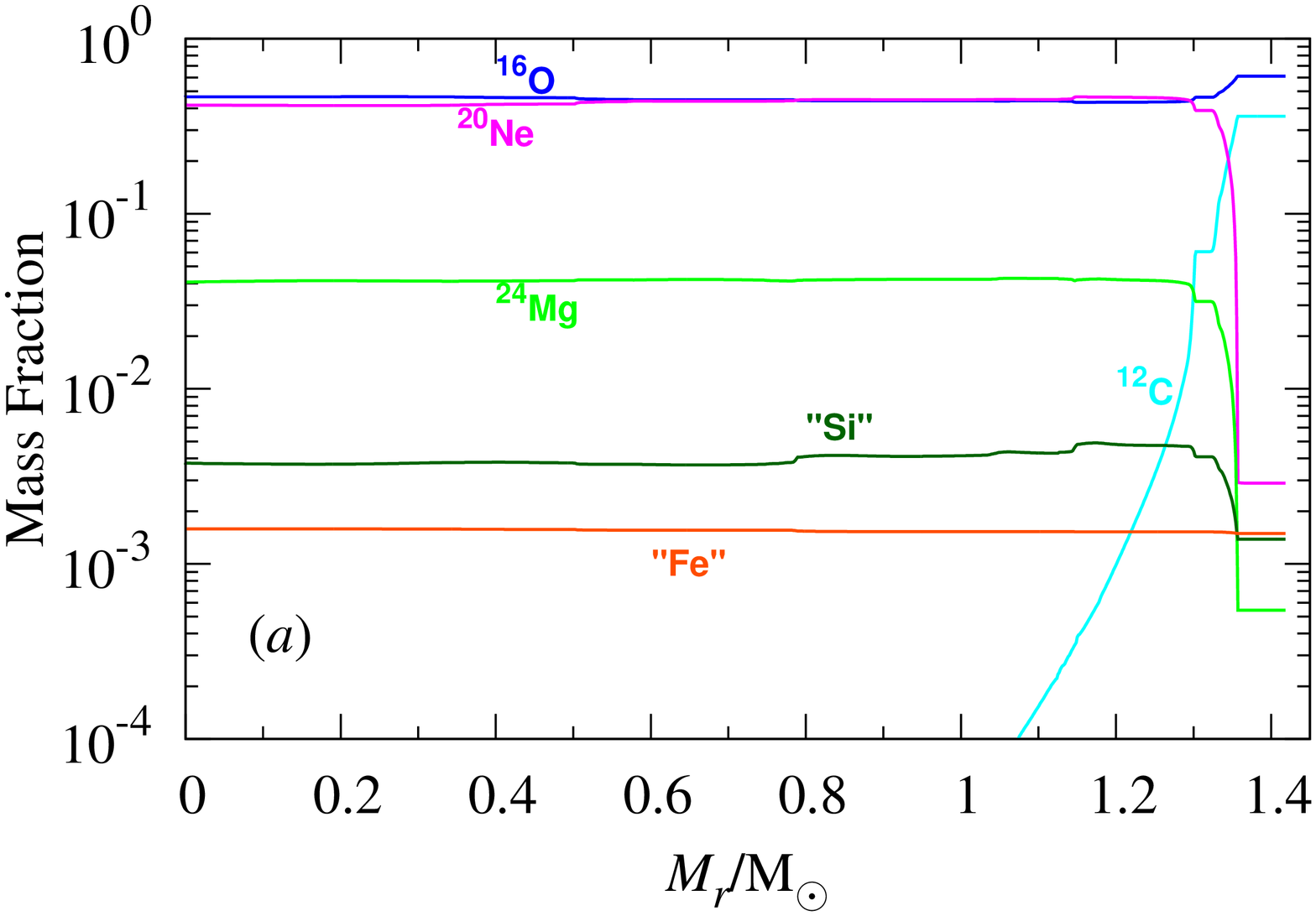}
\includegraphics[width=.2\textwidth,angle=270]{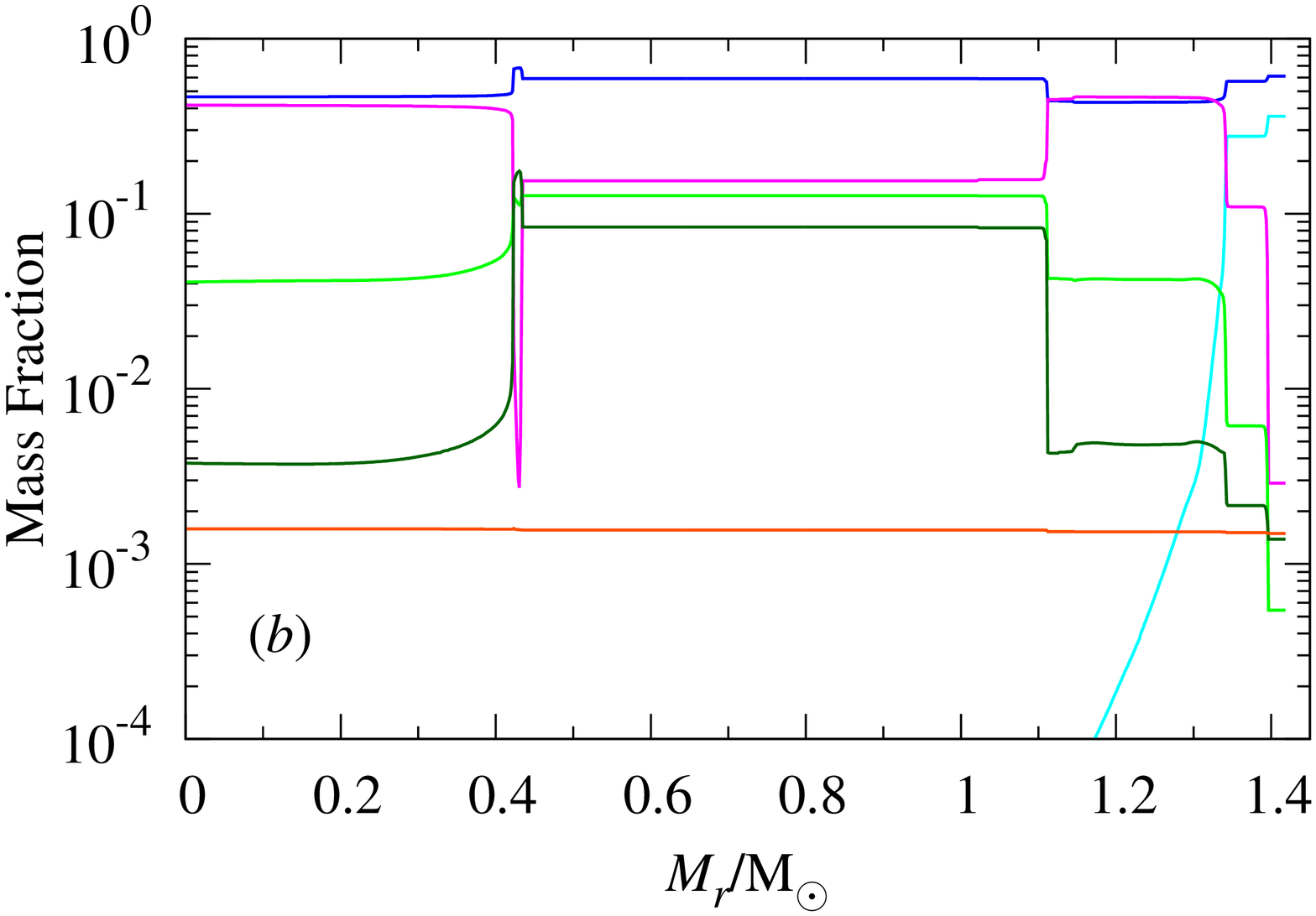}
\includegraphics[width=.2\textwidth,angle=270]{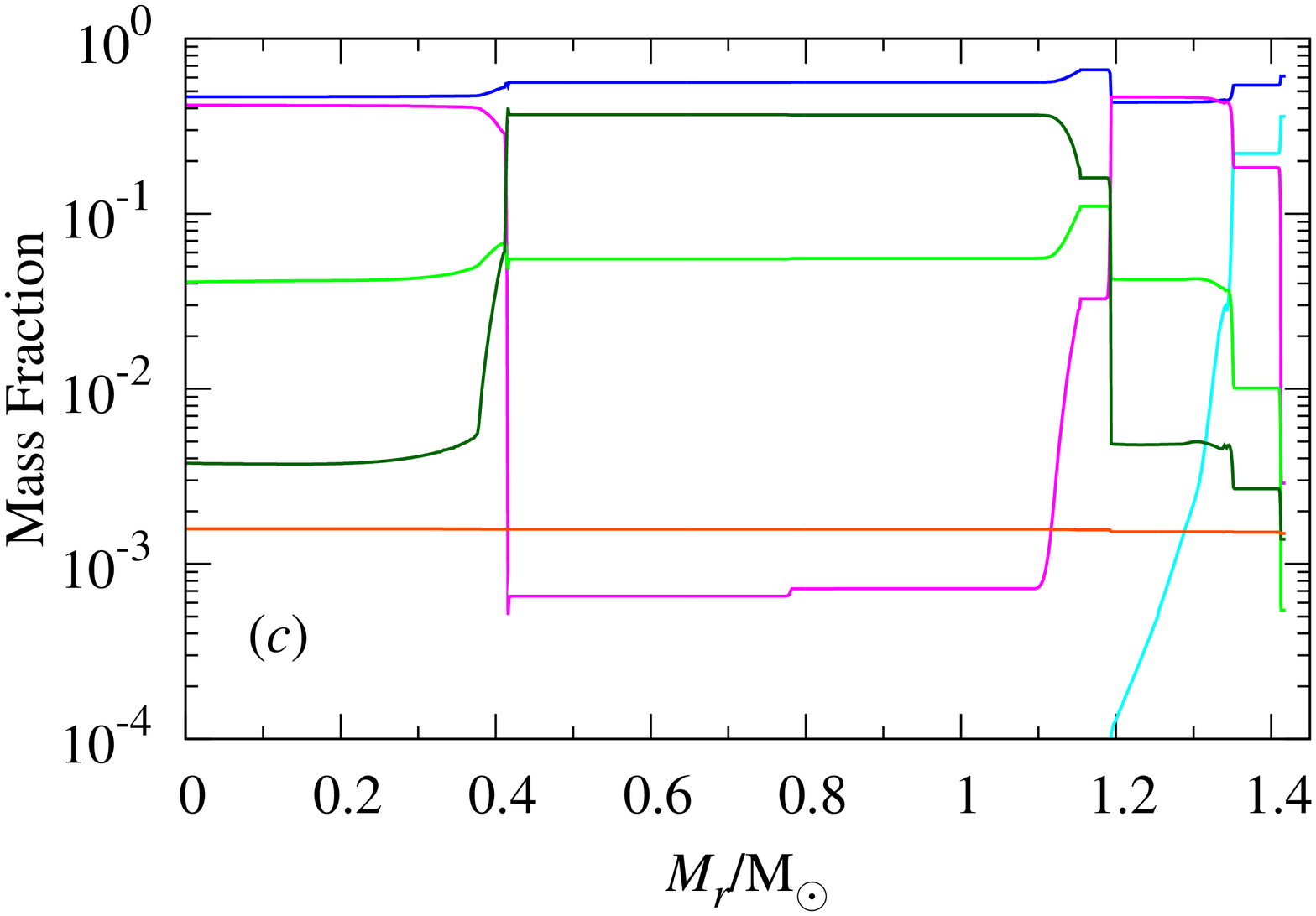}
\includegraphics[width=.2\textwidth,angle=270]{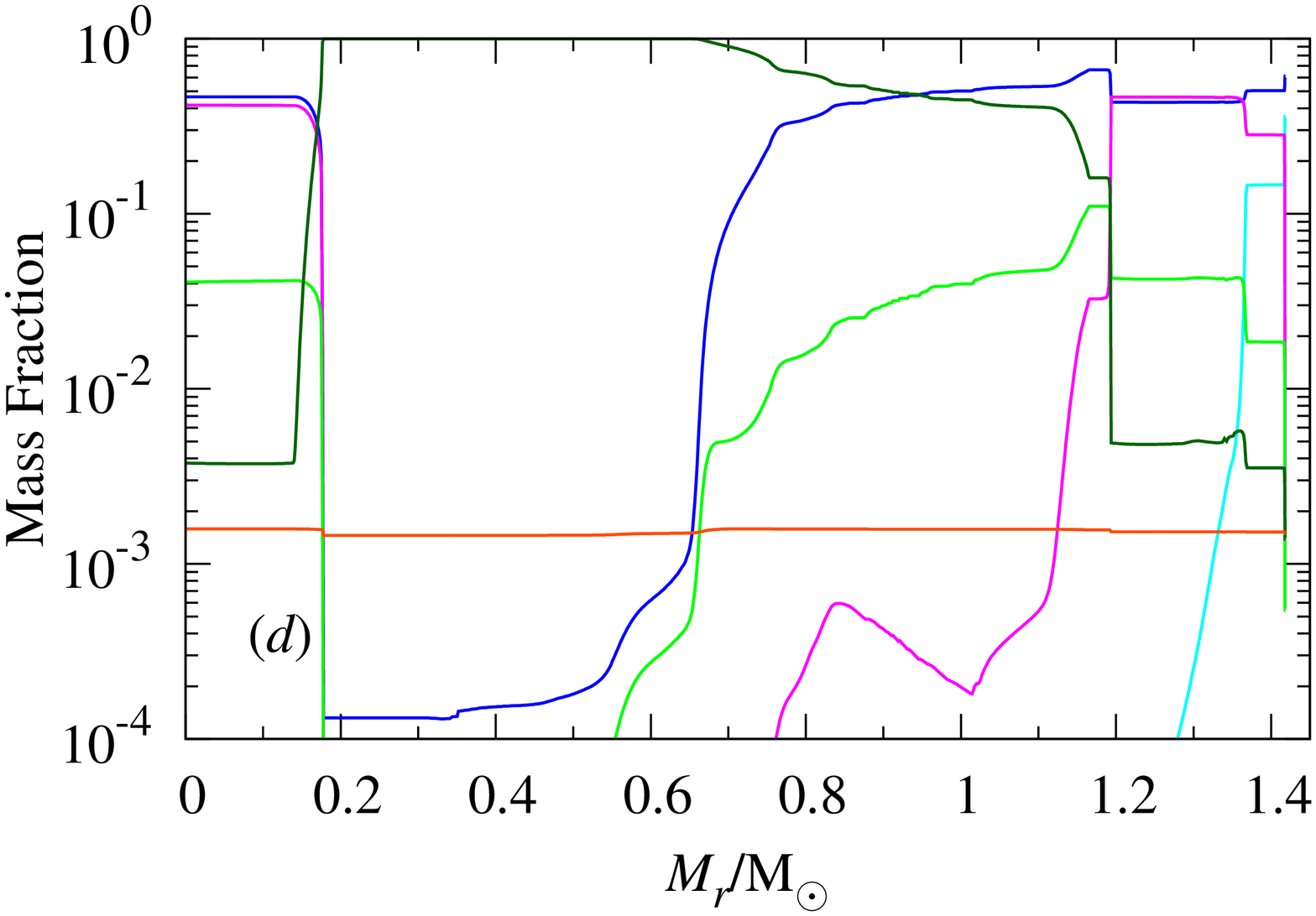}
\includegraphics[width=.2\textwidth,angle=270]{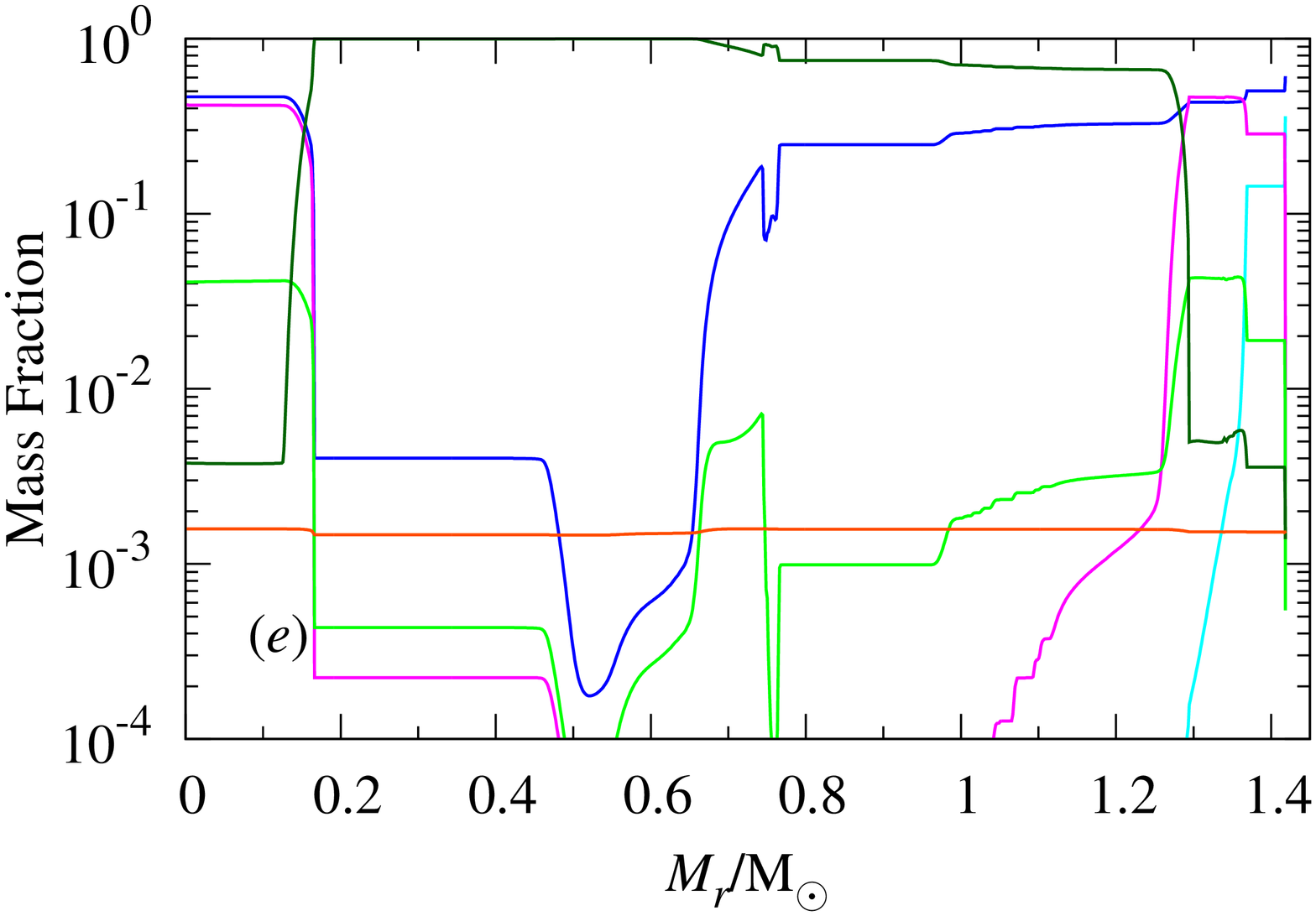}
\includegraphics[width=.2\textwidth,angle=270]{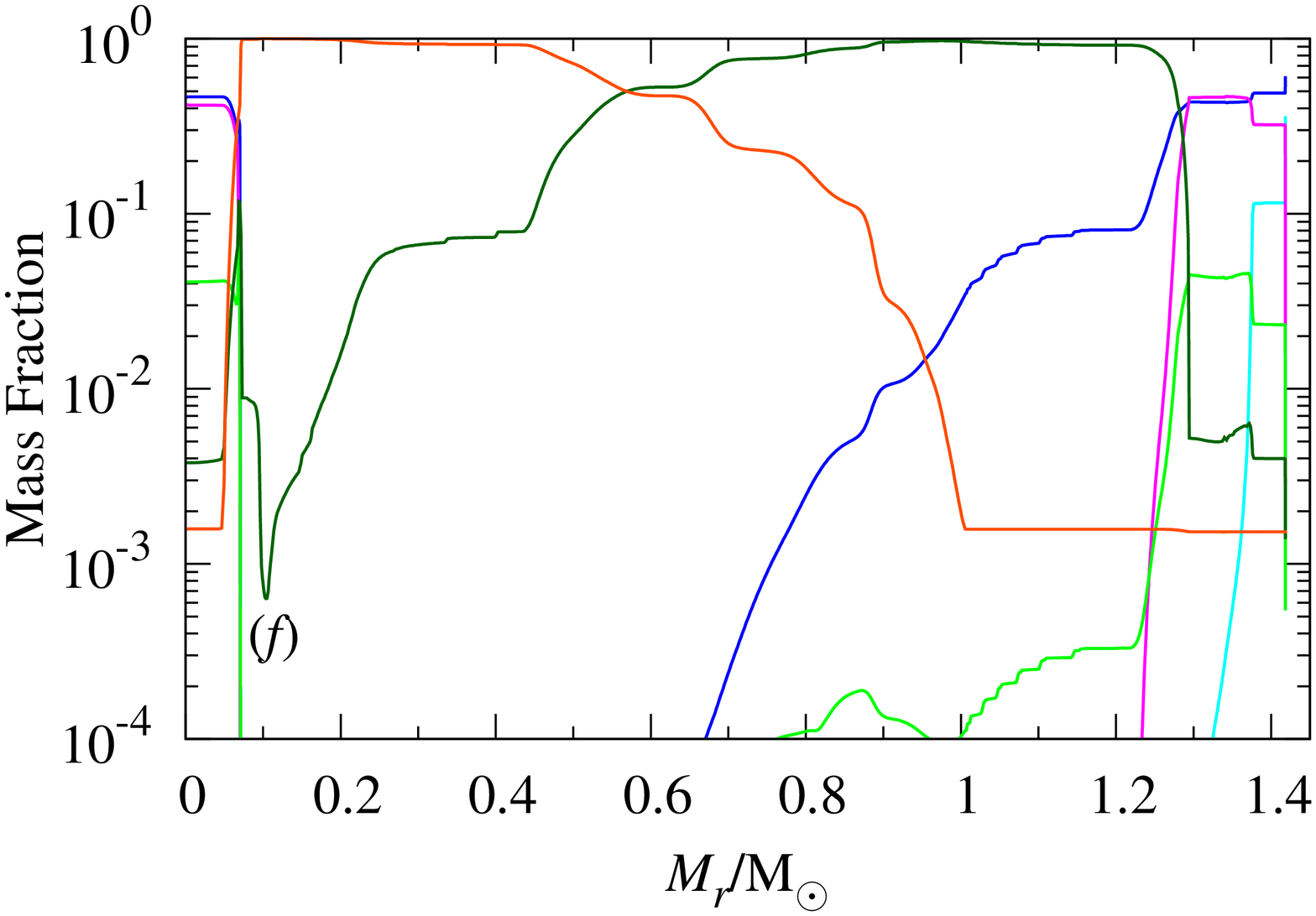}
\includegraphics[width=.2\textwidth,angle=270]{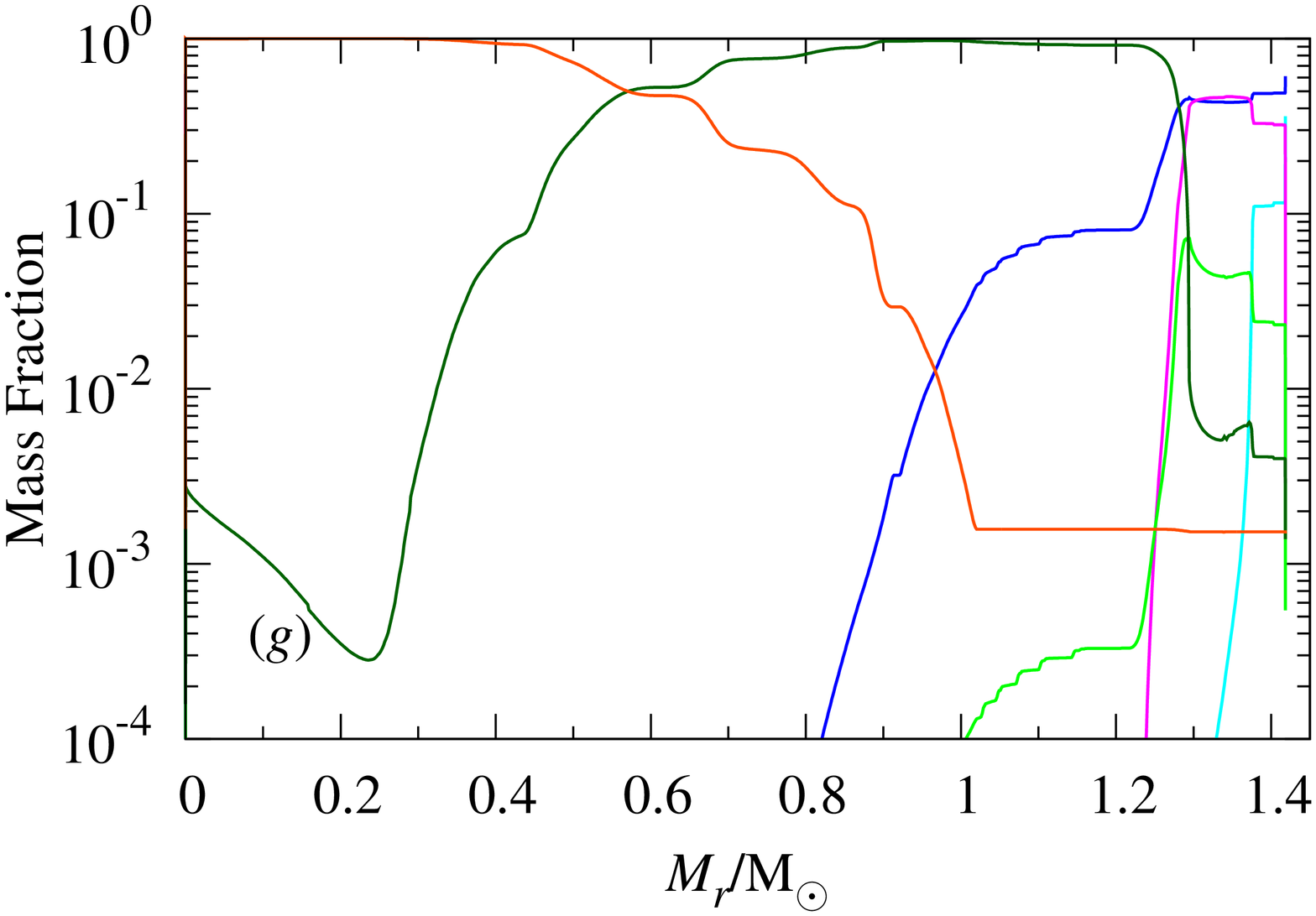}
\includegraphics[width=.2\textwidth,angle=270]{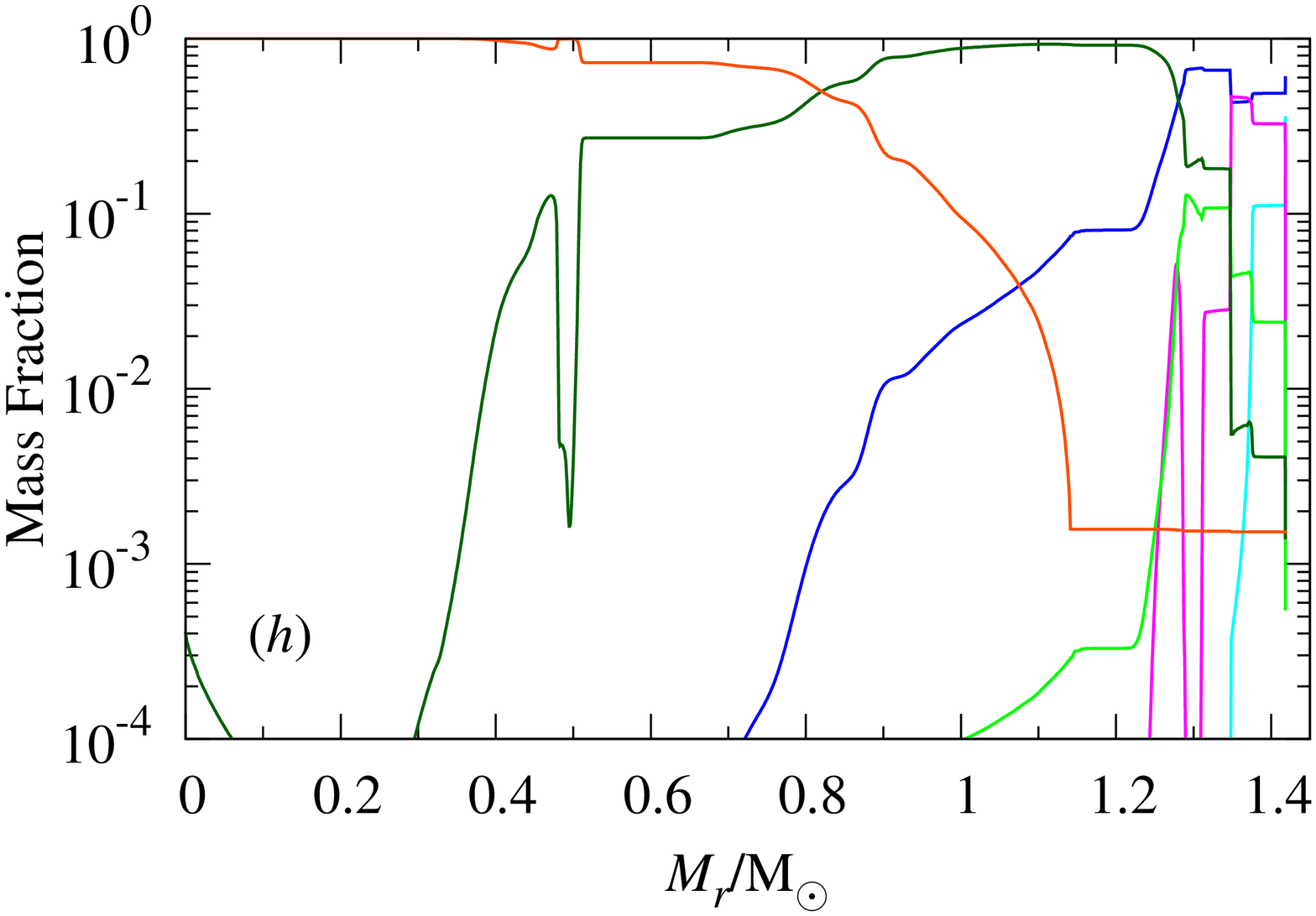}
\includegraphics[width=.2\textwidth,angle=270]{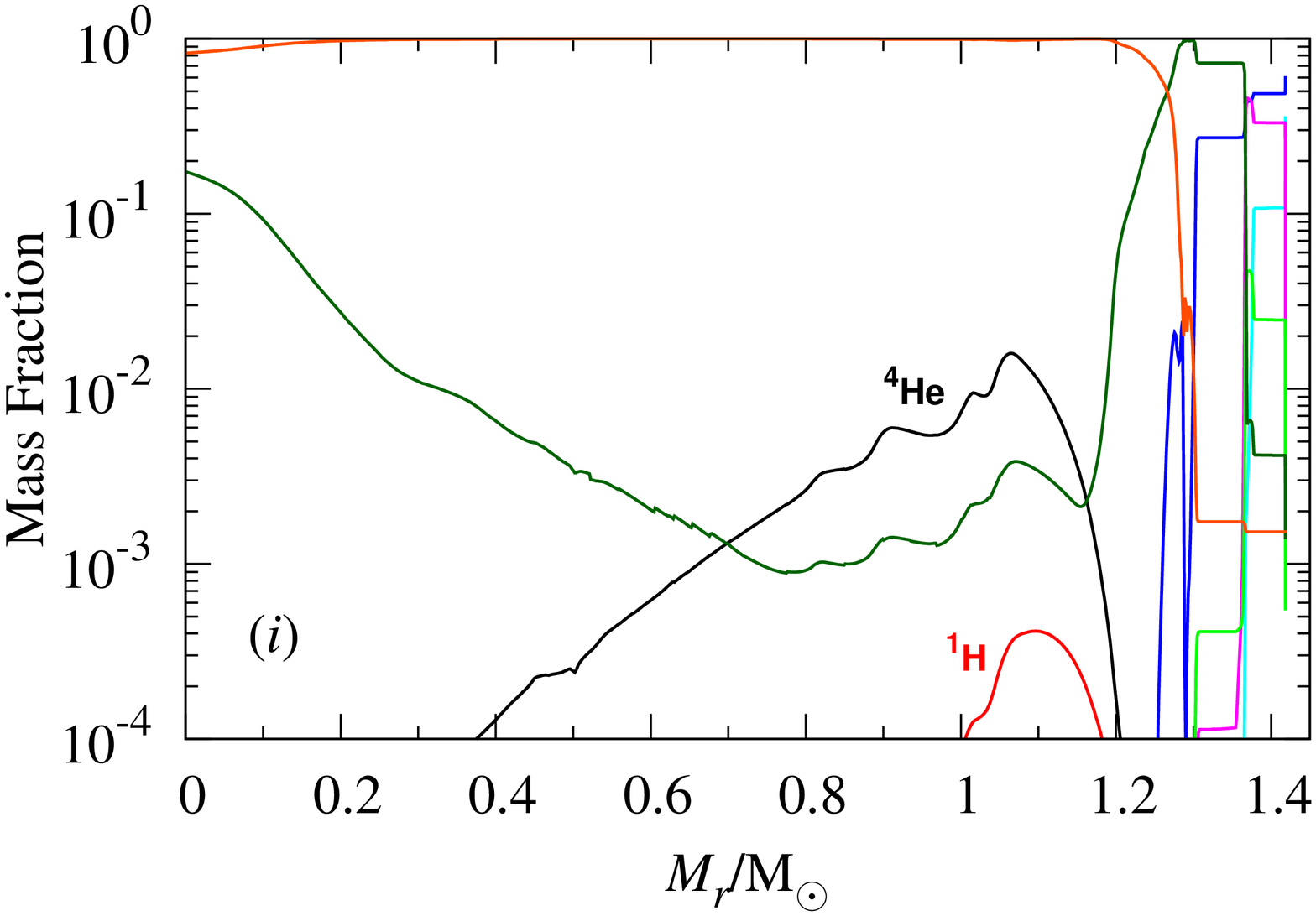}
\caption{Mass fraction distributions in different evolution
stages of CO142 model.
Panels ($a$)--($i$) correspond to the evolution number (2)--(10)
in Fig. \ref{fig:temp1p42co}.
}
\label{fig:mf142co}
\end{figure*}

We present evolution properties of light CO cores in which an off-center Ne ignition occurs.
Here we show the evolution of CO142 model as an example.
Figure \ref{fig:temp1p42co} shows the evolution of the temperature profile of CO142 model.
Figure \ref{fig:mf142co} shows the mass fraction distributions of nine different evolution stages corresponding to the lines 2--10 shown in Fig. \ref{fig:temp1p42co}.

Carbon ignites at the center for all of the models.
An ONe core (a central region with the C mass fraction being less than 0.01) forms through convective core C burning.
The line 1 of Fig. \ref{fig:temp1p42co} shows the temperature profile at $t_f - t = 1.0 \times 10^4$ yr for CO142 model, where $t_f$ is the time at the last step of the calculation.
The convective core C burning proceeds for $6.4 \times 10^3$ yr.
The core grows up through the following several C shell burnings.
The ONe core gradually contracts and the location of the maximum temperature moves outward.
The plasma neutrino process dominates in the neutrino energy loss in the temperature inversion region.
The line 2 in Fig. \ref{fig:temp1p42co} shows the temperature profile after the last C shell burning.
Figure \ref{fig:mf142co}($a$) shows the mass fraction distribution at that time.

Neon ignites at an off-center region after the ONe core mass becomes 1.32--1.33 M$_\odot$.
The burning front is formed at the inner edge of the burning layer.
The temperature at the burning front becomes the highest after the ignition, while the central temperature decreases (the line 3 of Fig. \ref{fig:temp1p42co} for CO142 model).
The off-center Ne burning makes a large convective O/Si layer on the ONe core (Fig. \ref{fig:mf142co}($c$) for CO142 model).
The convective layer extends to $\ge$ 1 M$_\odot$ in the mass coordinate. After the off-center Ne burning, the temperature at the inner edge of the O/Si layer decreases.
The ONe core contracts again.

Off-center O burning starts at the inner edge of the O/Si layer.
An inner region of the O/Si layer becomes the Si layer and a part of oxygen is burned in the other region.
Then, the inner edge of the Si layer gradually moves inward and, thus, the Si layer grows up inward.
The energy released in the burning front partially transfers into the core and the oxygen and neon in contact with the burning front are burned into silicon.
The line 5 of Fig. \ref{fig:temp1p42co} and Fig. \ref{fig:mf142co}($d$) show the temperature profile and the mass fraction distribution during the off-center O burning for CO142 model.
The mass of the ONe core is $\sim$0.2 M$_\odot$.
The Si-rich layer extends to $\sim$0.9 M$_\odot$, the O/Si layer ranges to $\sim$ 1.2 M$_\odot$, and the O/Ne layer remains on the O/Si layer.

During the off-center Ne and O burnings, the electron fraction in the O/Si and Si layers on the burning front is reduced through electron captures and $\beta^+$ decays, respectively.
The main parent nuclei for electron captures and $\beta^+$ decays are $^{26}$Al (isomeric state), $^{31}$S, and $^{30}$P.
The increase in the temperature at the burning front enhances these reactions.
During the off-center Ne burning of CO142 model, the electron fractions of the ONe core and the O/Si layer are 0.498 and 0.496, respectively.
The electron fraction in the O/Si layer becomes smaller than that in the ONe core.
Electron captures and $\beta^{+}$ decays at the burning front and the electron captures of $^{33}$S and $^{35}$Cl in the Si layer reduce the electron fraction in the Si layer.
The electron fraction in $M_r \sim 0.2$--0.5 M$_\odot$ 
reduces to $Y_e = 0.474$ during the off-center O burning for CO142 model.

Strong O shell burning occurs after the off-center O burning.
For CO142 model, the O shell burning extends the region of the Si layer up to $\sim$1.3 M$_\odot$ (Fig. \ref{fig:mf142co}($e$)).
This shell burning suppresses the core contraction and the inward motion of the inner edge for a while (the line 6 of Fig. \ref{fig:temp1p42co} for temperature proflie).

The inner edge of the Si layer continues to move inward and the temperature at the burning front gradually increases.
The intermediate elements outside the front are gradually burned to Fe-peak elements.
Then, the inner region of the Si layer changes to the Fe layer.
The electron fraction in this layer is reduced through electron captures of $^{54}$Mn, $^{55}$Mn, and $^{57}$Fe.
The burning front of CO142 model comes to $\sim$0.08 M$_\odot$ and the Fe layer grows up to 0.08--0.6 M$_\odot$ at $t_f - t = 4.7$ d (see the line 7 in Fig \ref{fig:temp1p42co} for the temperature profile and Fig. \ref{fig:mf142co}($f$) for mass fraction distribution).
The electron fraction at the inner edge of the Fe layer becomes $\sim$0.453.
The burning front moves inward further and it reaches the center
(see the line 8 of Fig. \ref{fig:temp1p42co} and Fig. \ref{fig:mf142co}($g$) for CO142 model).
At that time, the central temperature is the highest in the star.
The Fe ^^ ^^ core" of CO142 model is about 0.6 M$_\odot$ and thick Si-rich layer up to $\sim$1.3 M$_\odot$ is on the Fe core.

Then, the entire core contracts and the Fe core grows up through several Si shell burnings.
The line 9 of Fig. \ref{fig:temp1p42co} and Fig. \ref{fig:mf142co}($h$) show the temperature profile and the mass fraction distribution of CO142 model.
During the Si shell burnings, the shell-burning front sometimes becomes the highest temperature.
At this time, the Fe core grows up to $\sim$0.8 M$_\odot$.
The Fe core continues growing until the calculation is terminated.
The line 10 of Fig. \ref{fig:temp1p42co} and Fig. \ref{fig:mf142co}($i$)) shows the temperature distribution and the mass fraction distribution at the last step of CO142 model.
The Fe core has an approximately isothermal structure.
The Fe core is surrounded by thin Si, Si/O, O/Ne, and O/C layers.

\section{Modified Chandrasekhar limit}
\label{sec:app2}

In this section, we give a brief explanation of modified Chandrasekhar mass. From the hydrostatic equation,
\begin{equation}
\frac{dP}{dr}=-\frac{GM}{r^2}\rho,
\end{equation}
where $P$ is pressure, $r$ is radius, $G$ is the gravitational constant, $M$ is enclosed mass inside $r$, and $\rho$ is density, the dimensionless equation (so-called {\it Lane-Emden equation}) is derived as \citep[e.g.][]{shap83}
\begin{equation}
\frac{1}{\xi^2}\frac{d}{d\xi}\left(\xi^2\frac{d\theta}{d\xi}\right)=-\theta^N.
\label{eq:lane-emden}
\end{equation}
Here, the following equations are used:
\begin{align}
\rho&=\rho_c\theta(\xi)^N,\\
P&=P_c\theta(\xi)^{N+1},\\
r&=\alpha\xi,\\
\alpha&=\left(\frac{N+1}{4\pi G}\frac{P_c}{\rho_c^2}\right)^{1/2},
\end{align}
where $\rho_c$ and $P_c$ are the central density and pressure, and $N$ is the polytropic index. 
They are related to each other as $P_c=K\rho_c^{1+1/N}$ with a constant $K$.
By integrating Eq. \eqref{eq:lane-emden} for $\theta(\xi)$ from the center ($\xi=0$) toward the stellar surface ($\xi_N$ where $\theta(\xi_N)=0$) with boundary conditions, $\theta(0)=1$ and $\theta'(0)=1$, the stellar structure is determined.  From the solution the stellar radius and mass are given by
\begin{align}
R&=\alpha\xi_N=\left(\frac{N+1}{4\pi G}\frac{P_c}{\rho_c^2}\right)^{1/2}\xi_N,\\
M&=\int_0^R 4\pi\rho r^2dr=4\pi\alpha^3\rho_c\int_0^{\xi_N}d\xi \xi^2\theta^N\nonumber\\
&=-4\pi\alpha^3\rho_c\left.\left(\xi^2\frac{d\theta}{d\xi}\right)\right|_{\xi=\xi_N}.
\end{align}
By making use of
\begin{equation}
\varphi_N=-(N+1)^{3/2}\left.\left(\xi^2\frac{d\theta}{d\xi}\right)\right|_{\xi=\xi_N},
\end{equation}
the mass is given by
\begin{equation}
M=\left(\frac{1}{4\pi G^3}\frac{P_c^3}{\rho_c^4}\right)^{1/2}\varphi_N.
\label{eq:M}
\end{equation}
$\varphi_N$ depends on $N$. For instance, $N=3$ gives 
16.15 and $N=3.3$ gives 17.27. The equation of state for relativistic electrons is given by $P=K(\rho Y_e)^{4/3}$, where $Y_e$ is the electron fraction and $K=\frac{(3\pi^2)^{1/3}}{4}\frac{\hbar c}{m_N^{4/3}}$ with $\hbar$, $c$ and $m_N$ being reduced Planck constant, speed of light and nucleon mass, respectively. If $Y_e$ is constant, it corresponds to $N=3$. However, during the stellar evolution, $Y_e$ decreases due to the electron capture. The central value is approximately $Y_{e,c}\sim 0.42$ when $\rho_c\sim 10^{10}$ g cm$^{-3}$. Assuming $Y_{e,c}\propto \rho^\alpha$ and $Y_{e,c}=0.5$ for $\rho_c=10^7$ g cm$^{-3}$, which indicates $\alpha\approx -0.025$, we get $P_c\propto \rho_c^{\frac{4}{3}(1+\alpha)}\sim \rho_c^{1+\frac{1}{3.3}}$, thus $N\sim 3.3$.

In addition to the degeneracy pressure of electrons, the finite temperature correction is also important for the presupernova cores. Thus, the pressure is given by \citep{baro90}
\begin{equation}
P=K(\rho Y_e)^{4/3}\left[1+\frac{2}{3}\left(\frac{s_e}{\pi Y_e}\right)^2\right],
\label{eq:P}
\end{equation}
where $s_e$ is the electronic entropy. 

Combining Eqs. (\ref{eq:M}) and (\ref{eq:P}), we get
\begin{align}
M&=1.09M_\odot
\left(\frac{Y_{e,c}}{0.42}\right)^2
\left[1+\frac{2}{3}\left(\frac{s_{e,c}}{\pi Y_{e,c}}\right)^2\right]^{3/2}\nonumber\\
&\approx 
1.09M_\odot
\left(\frac{Y_{e,c}}{0.42}\right)^2
\left[1+\left(\frac{s_{e,c}}{\pi Y_{e,c}}\right)^2\right],
\end{align}
where $Y_{e,c}$ and $s_{e,c}$ are the central values of $Y_e$ and $s_e$. Here we use $\varphi_N=17.27$ (corresponding to $N=3.3$), since we are now interested in the stellar structure in which the electron capture reduces $Y_e$. We also assume that the thermal correction (the second term in the square brackets) is much smaller than unity.



\bsp	
\label{lastpage}
\end{document}